\documentclass[conference]{IEEEtran}
\IEEEoverridecommandlockouts
\usepackage{cite}
\usepackage{amsmath,amssymb,amsfonts}
\usepackage{algorithmic}
\usepackage{graphicx}
\usepackage{textcomp}
\usepackage{xcolor}
\usepackage{booktabs}
\usepackage{multirow}
\usepackage{makecell}
\usepackage{bm}
\def\BibTeX{{\rm B\kern-.05em{\sc i\kern-.025em b}\kern-.08em
    T\kern-.1667em\lower.7ex\hbox{E}\kern-.125emX}}
\begin{document}

\title{Self-Supervised Learning for User Localization}

\author{
\IEEEauthorblockN{Ankan Dash\textsuperscript{a}, Jingyi Gu\textsuperscript{a}, Guiling Wang\textsuperscript{a}, Nirwan Ansari\textsuperscript{b}}
\IEEEauthorblockA{\textsuperscript{a}\textit{Department of Computer Science} \\
\textit{New Jersey Institute of Technology}\\
Newark, United States \\
\{ad892@njit.edu, jg95@njit.edu, gwang@njit.edu\}}
\IEEEauthorblockA{\textsuperscript{b}\textit{Department of Electrical and Computer Engineering} \\
\textit{New Jersey Institute of Technology}\\
Newark, United States \\
nirwan.ansari@njit.edu}
}

% \and
% \IEEEauthorblockN{4\textsuperscript{th} Given Name Surname}
% \IEEEauthorblockA{\textit{dept. name of organization (of Aff.)} \\
% \textit{name of organization (of Aff.)}\\
% City, Country \\
% email address or ORCID}

\maketitle

\begin{abstract}

Machine learning techniques have shown remarkable accuracy in localization tasks, but their dependency on vast amounts of labeled data, particularly Channel State Information (CSI) and corresponding coordinates, remains a bottleneck. Self-supervised learning techniques alleviate the need for labeled data, a potential that remains largely untapped and underexplored in existing research. Addressing this gap, we propose a pioneering approach that leverages self-supervised pretraining on unlabeled data to boost the performance of supervised learning for user localization based on CSI. We introduce two pretraining Auto Encoder (AE) models employing Multi Layer Perceptrons (MLPs) and Convolutional Neural Networks (CNNs) to glean representations from unlabeled data via self-supervised learning. Following this, we utilize the encoder portion of the AE models to extract relevant features from labeled data, and finetune an MLP-based Position Estimation Model to accurately deduce user locations. Our experimentation on the CTW-2020 dataset, which features a substantial volume of unlabeled data but limited labeled samples, demonstrates the viability of our approach. Notably, the dataset covers a vast area spanning over $646 \times 943 \times 41$
meters, and our approach demonstrates promising results even for such expansive localization tasks. 

\end{abstract}


\section{Introduction}
%User localization plays a pivotal role in the realm of wireless communication systems, enabling an array of applications spanning navigation, smart factories, surveillance, security, and the Internet of Things (IoT). Accurate positioning not only benefits end-users but also enhances critical aspects of wireless technology, including radio resource management, beamforming, and channel estimation. There have been previous works which use deep learning to perform user localization\cite{8446013,8662548,Cerar2021ImprovingCM,9838535,8390298,Li_2019, 8711803, 9128759, 7127718, Sobehy2022}. However, the prevalent machine learning-based localization methods, while capable of achieving high accuracy, pose a formidable challenge in terms of data acquisition. These approaches necessitate vast amounts of labeled data, particularly Channel State Information (CSI) paired with corresponding coordinates. 

User localization is crucial in the domain of wireless communication systems, enabling a broad spectrum of applications including navigation, smart factories, surveillance, security, and the Internet of Things (IoT) \cite{8889728}. Accurate positioning not only augments user experience but also bolsters vital aspects of wireless technology such as radio resource management, beamforming, and channel estimation. Existing work has leveraged deep learning to perform user localization\cite{8446013,8662548,Cerar2021ImprovingCM,9838535,8390298,Li_2019, 8711803, 9128759, 7127718, Sobehy2022}. However, the prevalent machine learning-based localization methods, while adept at achieving high accuracy, encounter significant hurdles in data acquisition. Specifically, these methods require substantial quantities of labeled data, particularly Channel State Information (CSI) paired with corresponding coordinates.

%The advent of self-supervised learning, fortunately, brings a transformative potential to the field of user localization. Self-supervised learning algorithms extract valuable features and patterns from data such as CSI measurements to create rich and context-aware embeddings. These learned representations can encapsulate a deep understanding of the CSI features' inherent structure and semantics, such as multipath characteristics and signal variations in wireless environments, inherently containing rich spatial and temporal information that can be harnessed for location prediction. What makes self-supervised learning particularly compelling is its potential to serve as a pre-training step for supervised learning tasks. By transferring the knowledge encoded in these representations to downstream supervised models, self-supervised learning effectively equips them with a data-driven intuition, often resulting in improved performance, robustness, and generalization on tasks with limited labeled data. 

The emergence of self-supervised learning heralds transformative potential in the realm of user localization. Algorithms under this learning paradigm are adept at extracting valuable features and patterns from data, such as CSI measurements, to construct rich and context-aware embeddings. These learned representations encapsulate a profound understanding of the inherent structure and semantics of CSI features, such as multipath characteristics and signal variations within wireless environments. This encapsulation inherently embodies rich spatial and temporal information that can be harnessed for location prediction. The allure of self-supervised learning largely lies in its capability to serve as a pre-training step for supervised learning tasks. By transmuting the knowledge encapsulated in these representations to downstream supervised models, self-supervised learning substantially augments them with a data-driven intuition. This often translates to enhanced performance, robustness, and generalization, especially in tasks constrained by limited labeled data.

%The advent of self-supervised learning, fortunately, brings a transformative potential to the field of user localization. Self-supervised learning algorithms extract valuable features and patterns from data such as CSI measurements to create rich and context-aware embeddings. These learned representations can encapsulate a deep understanding of the CSI features' inherent structure and semantics, such as multipath characteristics and signal variations in wireless environments, inherently containing rich spatial and temporal information that can be harnessed for location prediction. What makes self-supervised learning particularly compelling is its potential to serve as a pre-training step for supervised learning tasks. By transferring the knowledge encoded in these representations to downstream supervised models, self-supervised learning effectively equips them with a data-driven intuition, often resulting in improved performance, robustness, and generalization on tasks with limited labeled data.  

%In the current research landscape, a significant gap exists, as the majority of studies heavily rely on labeled data for user localization. While acquiring CSI data is relatively straightforward, obtaining precise user location labels demands substantial investments, extensive resources, and significant time. To the best of our knowledge, no prior research has explored the untapped potential of utilizing extensive unlabeled data.

In the prevailing research landscape, there exists a notable gap, as the lion's share of studies predominantly relies on labeled data for user localization. Although the acquisition of CSI data is relatively straightforward, securing accurate user location labels necessitates extensive resources and substantial time. To the best of our knowledge, no prior research has explored the untapped potential of utilizing extensive unlabeled data.

Our paper aims to bridge this gap by introducing an innovative approach. We harness self-supervised learning techniques on unlabeled CSI data to enhance the performance of supervised learning models in predicting user locations. By uncovering latent patterns and representations within unlabeled CSI data, we aim to improve the generalization and robustness of supervised models, reducing the need for extensive labeled datasets and providing more reliable location predictions. This study underscores the synergistic potential of self-supervised learning and supervised learning, highlighting how the former can catalyze advancements in user location prediction.

Our contributions can be summarized as follows:
(1) We are the first to build a pretrain model to learn representations from CSI, enabling the full utilization of the large unlabeled data. 
(2) We develop four models based on supervised and self-supervised learning using Multi-Layer Perceptron (MLP) and Convolutional Neural Network (CNN). Experiment results show that pretraining in our approach significantly improves the user localization prediction.
(3) Our approach highlights the effectiveness of self-supervised learning using unlabeled data as a powerful tool to augment the performance of supervised learning when the labeled data is scarce. 

\section{Related Works}
Recent research has explored the possibility of utilizing massive multiple-input multiple-output (MIMO) CSI data for user localization due to its ability to provide rich spatial information and high resolution. In this section, we review some existing works that use deep learning techniques to infer user location coordinates based on CSI data.

One of the earliest works in this field was conducted by Arnold et al.\cite{8446013}. They proposed a deep learning based user localization method using massive MIMO CSI data. They reduced the required amount of measured training data by first training DNNs on simulated line of sight (LoS) data and finetuning on measured non-line of sight (NLoS) data. 
Cerar et al.\cite{9482604} focused on indoor positioning using CSI data. They leveraged CNNs to improve the accuracy of indoor positioning, a crucial aspect in applications like indoor navigation and tracking. 
Their work yielded mean errors between 2cm to 10cm across diverse scenarios, utilizing the CTW-2019 dataset that spans an area of 4m x 2m and contains approximately 17,486 labeled samples.
Additionally, Wu et al.\cite{8390298} proposed a DNN-based Fingerprinting (FP) system employing a singular DNN to learn the mapping from CSI measurements to receiver positions. They employed a stack of autoencoders to learn pretrained weights. In a related vein, Hsieh et al.\cite{8662548} used deep learning for indoor localization, segmenting a room into 2D blocks treated as classes. Using MLPs and 1D CNNs, they simplified location estimation by predicting a subject's presence in a block rather than precise coordinates.
Furthermore, Foliadis et al. \cite{9838535} employed Deep Learning on CSI fingerprints and multiple base stations to attain accurate localization in wireless networks. They proposed a two-stage localization methodology: initially predicting the user's position for each base station independently, then aggregating predictions to yield a more accurate and reliable localization estimate. Notably, the uncertainty in the User Equipment's (UE's) localization at each base station was factored in while aggregating the predictions.

From the summaries of previous research, it's clear that most of them depend on labeled data for user localization. Some even simplify it by using 2D block classification. However, collecting labeled data is time and resource-intensive. Hence, our work investigates using abundant unlabeled data for self-supervised learning to enhance supervised learning when labeled data is scarce. We demonstrate that self-supervised learning can significantly improve supervised learning performance in low-labeled data scenarios.

\section{Methodology}

In this section, we outline our methodology of employing self-supervised learning on a large unlabeled dataset for pretraining to generate representations of CSI features, followed by using supervised learning on a limited labeled dataset for finetuning, thereby enhancing user localization performance.

\subsection{Problem Formulation}

User localization is to determine a user's position precisely based on Channel State Information (CSI). Each data sample contains an estimated channel frequency response between the user $i$ and an antenna array, denoted as $\mathbf{x}_i\in\mathbb{R}^{a\times s \times m}$. Here, $a$ represents the number of working antennas, $s$ is the number of used subcarriers, and $m$ is the total number of measurements per location. Additionally, each data point is accompanied by a ground truth position $\mathbf{p}_i\in\mathbb{R}^3$ representing three dimensions in the Cartesian coordinate system. We aim to build a neural network capable of taking CSI features $\mathbf{x}_i$ as input and predicting the 3D position $\mathbf{y}_i\in\mathbb{R}^3$.

\subsection{Pretraining via Reconstruction}
In utilizing the extensive unlabeled dataset, we adopt a self-supervised learning approach \cite{balestriero2023cookbook} for pretraining, with the aim to learn representations of the CSI. The objective in our pretraining phase is set as the reconstruction of the CSI information. We utilize an autoencoder (AE) structure \cite{DBLP:journals/corr/abs-2003-05991}, parameterized by  $\theta$, to derive compact and informative representations from the CSI measurements, obviating the need for manual labeling.

The AE comprises two essential components: an encoder that takes CSI $\mathbf{x}$ as input and generates the latent representation $\mathbf{z}$, and a decoder that reconstructs the CSI $\mathbf{r}$ from this latent representation $\mathbf{z}$. This process enables the extraction of meaningful features. We train the model as a reconstructing AE by minimizing the difference between the original and reconstructed CSI. The loss with respect to $\theta_e$ and $\theta_d$ are presented as follows:

\begin{gather}
\mathcal{L}_p=\mathbb{E}[\left\| \mathbf{x}-\mathbf{r} \right\|]_2^2 \\
\mathbf{z}=Encoder_{\theta_e}(\mathbf{x}),\quad \mathbf{r}=Decoder_{\theta_d}(\mathbf{z})
\end{gather}

Specifically, we design two distinct types of architecture for AE model, based on MLPs and CNNs:

\subsubsection{MLP-based AE}
Our MLP-based encoder contains $k_m^e$ fully connected layers with ReLU as the activation function, and the decoder contains $k_m^d$ linear and ReLU layers.

\subsubsection{CNN-based AE}

Within CNN-based AE, the encoder contains $k_c^e$ convolutional layers followed by ReLU activation and max pooling layers, the decoder operates $k_c^d$ layers of 2D transposed convolution followed by ReLU activation. 

\subsection{Finetuning via Position Estimation Model}

Following the pretraining phase on the extensive unlabeled dataset, the pretraining model becomes adept at capturing meaningful representations from CSI features. We then shift to supervised learning \cite{Cunningham2008} for finetuning the model parameters for the downstream task, user localization, employing a limited labeled dataset, where the CSI measurements are paired with corresponding ground truth user locations. Through this pairing, the supervised model can learn the intricate relationship between the observed channel characteristics and the physical positions of users. 

Specifically, the pretrained encoder in the AE extracts latent representations $\mathbf{z}$ from CSI features $\mathbf{x}$ in the labeled data. These representations are then processed by an MLP-based position estimation model, which comprises a series of linear layers with ReLU activation to predict the user's location $\mathbf{y}$. The finetuning process is trained by minimizing the loss $\mathcal{L}_f$, which measures the discrepancy between ground truth and predicted 3D coordinates:

\begin{gather}
    \mathcal{L}_f = \mathbb{E}[\left\| \mathbf{p}-\mathbf{y} \right\|]_2^2, \quad
    \mathbf{y} = MLP(\mathbf{z})
\end{gather}

%Supervised learning is a foundational paradigm in machine learning and artificial intelligence, characterized by its capacity to teach computers to make predictions or classifications based on labeled data. (\textcolor{red}{this paragraph is about what is SSL and advantage, how about moving it into the introduction?}
\section{Experiments}

\subsection{Dataset}
\begin{figure}[h!]
    \centering
    \includegraphics[width=0.49\textwidth]{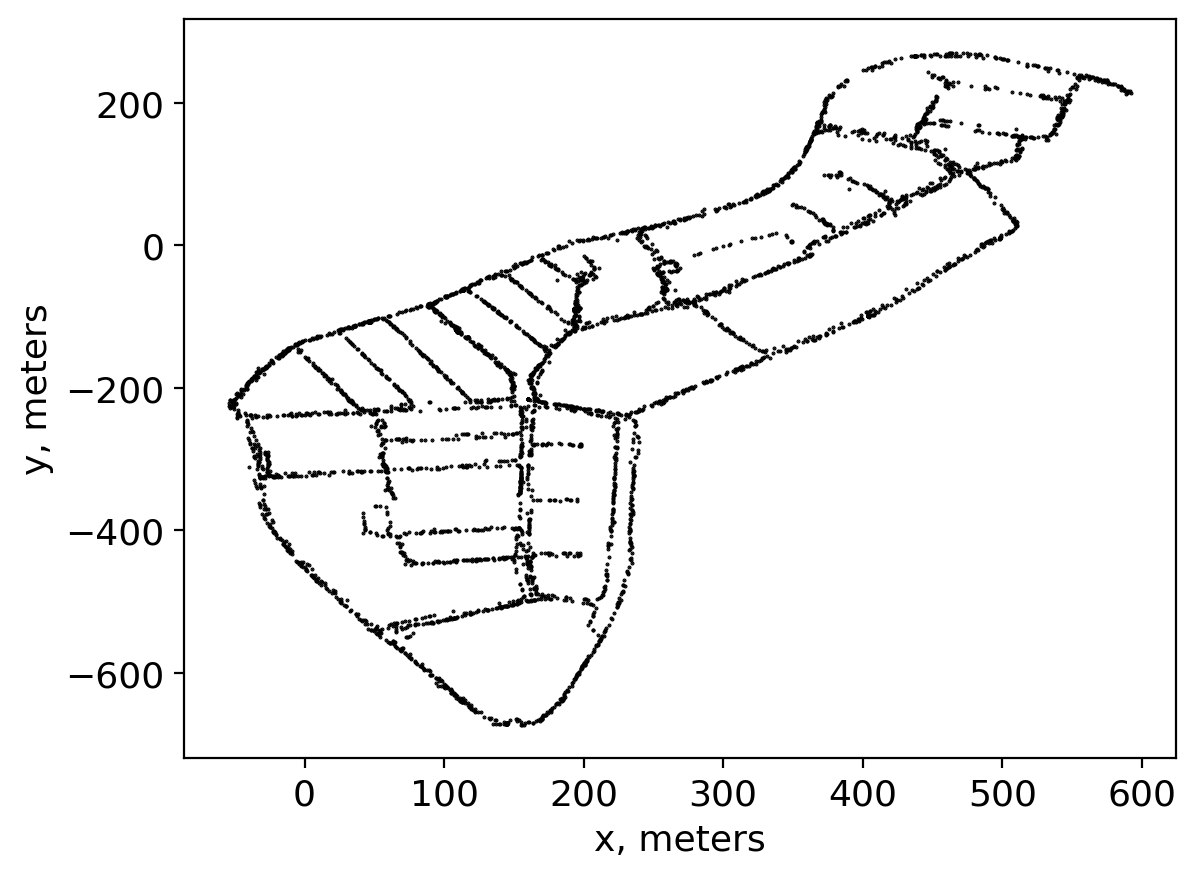}
    \caption{User Equipment position on the XY plane with dimensions in meters and base station at (0,0).}
    \label{fig:maparea}
\end{figure}

We utilize the IEEE CTW-2020 dataset available on IEEE Machine Learning for Communication website\cite{ieee_ctw-2020}. This dataset contains an unlabeled dataset and a labeled dataset, both of which can be downloaded from the same source. 
%Figure \ref{fig:maparea} shows the User Equipment positions in the XY plane with respect to the base station which is at (0,0). It covers a substantially large area of $646\times943\times41$ meters. 
Figure \ref{fig:maparea} illustrates the positions of the User Equipment (UE) in the XY plane relative to the base station situated at coordinates (0,0). The covered area spans a substantial dimension of $646\times943\times41$ meters.

The unlabeled data comprises a total of 36,192 samples, each containing the real part of estimated channel matrices as the CSI features. These matrices are structured with dimensions of $[56, 924, 5]$. 
The labeled data also has CSI features sharing the same dimension with unlabeled data. Besides, it augments this information with the target positions as the ground truth positions of the transmitter. It is represented in the Cartesian coordinate system, each sample has the shape of $[x, y, z]$. In both labeled and unlabeled data, CSI features are averaged over the last dimension because they are basically five measurements for a single data sample. After taking the mean, each sample in both datasets has a shape of $[56, 924]$.

During the pretraining phase, the unlabeled data is randomly partitioned into training and validation sets in an 8:2 ratio. The model weights yielding the minimal validation loss are preserved. During the finetuning phase, the labeled data is randomly divided into the training, validation, and test datasets with a ratio of 90:5:5. Hyperparameters tuning is performed based on validation data. Accordingly, we save the best model weights and then test them on the test dataset to evaluate their performance.

\subsection{Baselines}
To thoroughly evaluate the impact of unlabeled data, the potential of pretraining, and the influence of different model structures, we carry out a set of four experiments, each serving a distinct purpose:

\subsubsection{Supervised learning using labeled data only}

\begin{figure}[h!]
    \centering
    \includegraphics[width=0.49\textwidth]{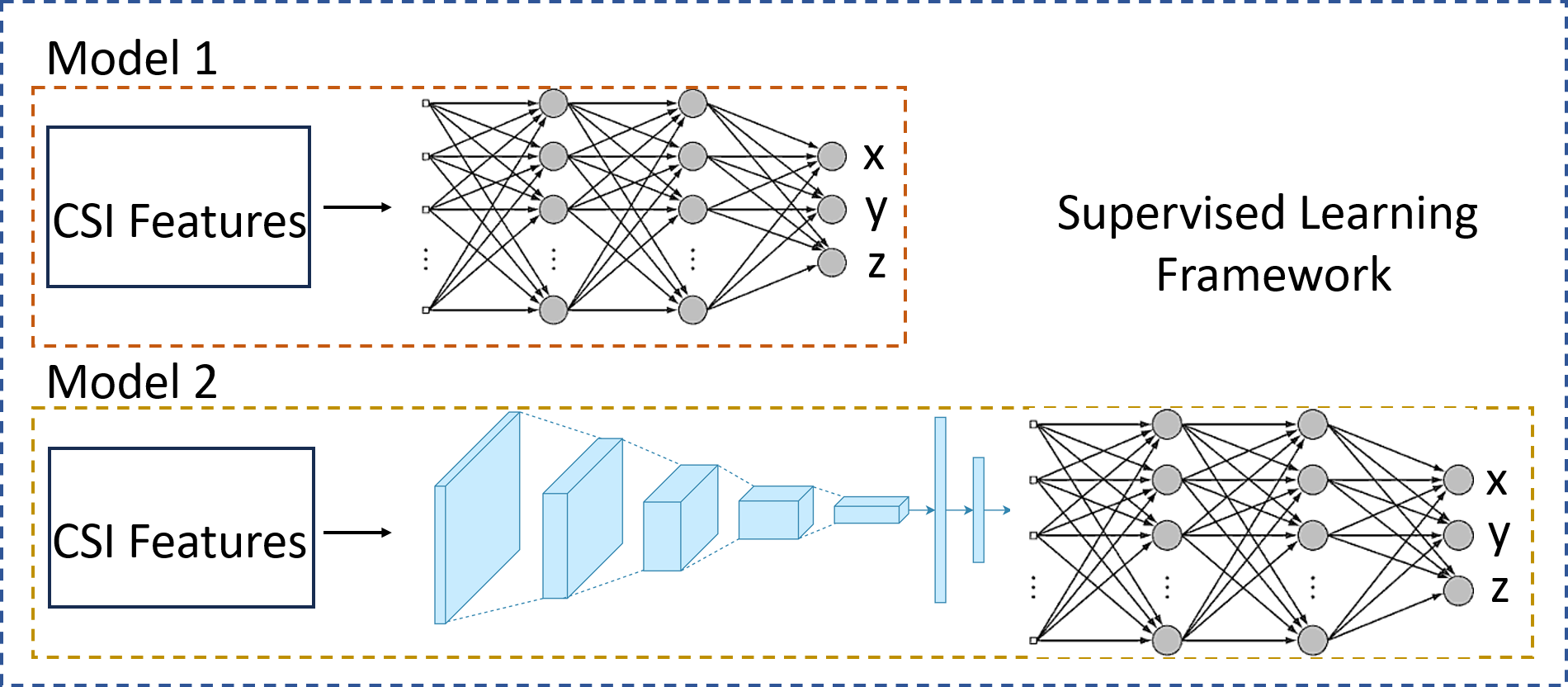}
    \caption{Model 1 and Model 2 architecture for Supervised Learning with labeled data.}
    \label{fig:superVised}
\end{figure}

In this category, we employ two models for supervised learning, using labeled data only to directly predict the user's $x$, $y$, and $z$ axis by learning from the available CSI features. Figure \ref{fig:superVised}. shows the two supervised models that we use. 

\begin{itemize}
    \item \textbf{Model 1} is an MLP-based neural network model. It contains 3 linear layers with hidden dimensions of 128, 64, and 3 and ReLU activation.
    \item \textbf{Model 2} is a CNN-based model. It contains 2 layers of CNN with kernel sizes of 3 and 2 and hidden dimensions of 32 and 64, and ReLU, max pooling, and linear layers. The architecture of CNN can capture spatial information from coordinates.
\end{itemize}

\subsubsection{Pretraining on unlabeled data and finetuning using labeled data}

\begin{figure}[h!]
    \centering
    \includegraphics[width=0.49\textwidth]{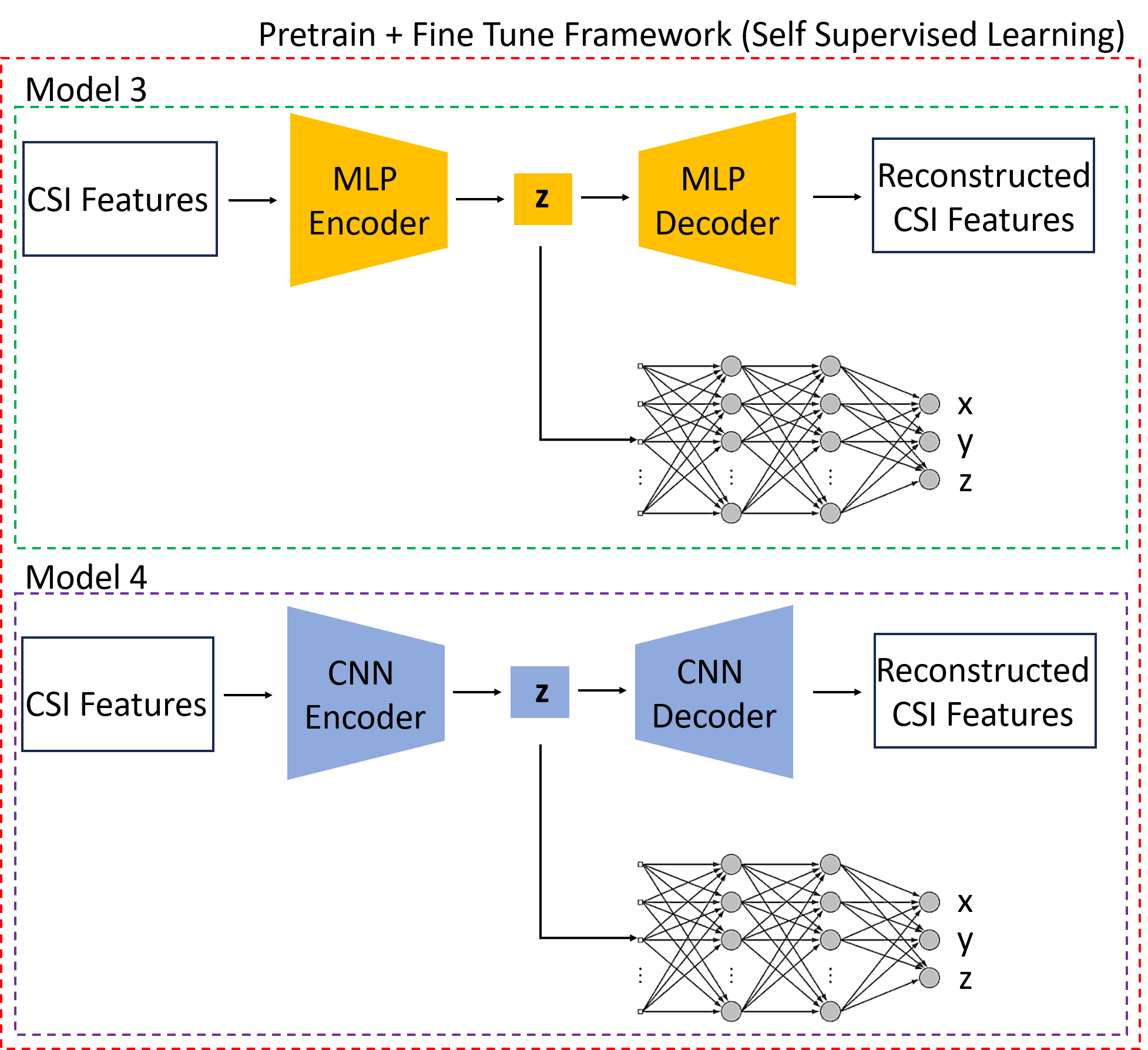}
    \caption{Model 3 and Model 4 architecture for pretraining and finetuning with unlabeled and labeled data.}
    \label{fig:selfsupvised}
\end{figure}

In this category, two models first leverage unlabeled data for pretraining via unsupervised learning and then use the encoder of the pretraining model to extract features from the labeled data. These features are then given as input to an MLP-based position estimation model to predict the user location via supervised learning. Figure \ref{fig:selfsupvised}. shows the self-supervised framework which uses pretraining with unlabeled data and finetuning with labeled data.

\begin{itemize}
    \item \textbf{Model 3} employs an MLP-based AE model in the pretraining phase to learn the informative representations from the unlabeled data. The encoder consists of 4 linear layers with hidden dimensions of 256, 128, 64, and 32 as well as ReLU. The decoder consists of 4 linear layers in the opposite order.
    \item \textbf{Model 4} utilizes a CNN-based AE pretraining model to extract the CSI feature representations. The encoder contains 2 convolutional layers with hidden dimensions of 32 and 64, kernel sizes of 3 and 2, followed by ReLU and MaxPooling layers. The decoder includes 2 convolution transpose layers with hidden dimensions of 32 and 1, and kernel sizes of 3, followed by ReLU.
\end{itemize}

The structure of position estimation models in Model 3 and 4 during the finetuning phase are the same as in Model 1.

All models are trained on a Nvidia Titan RTX. The batch size is set to 64. We choose Adam optimizer with a learning rate of 0.001. Models are trained for 100 epochs with early stopping.

\subsection{Evaluation metrics}
The performance of the proposed models is evaluated by the following metrics. Each metric is computed in three dimensions within the Cartesian System. To fully evaluate the performance, we also record the metrics that are averaged across all three dimensions:

\begin{enumerate}
    \item \textbf{Mean absolute error (MAE)} measures the absolute error between predicted and true positions.
    \begin{gather}
        MAE_m = \frac{1}{n}\sum_{i}^{n}{|p_{i,m} - y_{i,m}|} \\
        MAE_a = \frac{1}{3n}\sum_{m}^3\sum_{i}^{n}{|p_{i,m} - y_{i,m}|} 
    \end{gather}
    \item \textbf{Normalized mean absolute error (NMAE)} calculates the mean absolute error averaged by the range of coordinates in each axis.
    \begin{gather}
        NMAE_m = \frac{1}{n}\sum_{i}^{n}{|\frac{p_{i,m} - y_{i,m}}{max(\mathbf{p}_{m})-min(\mathbf{p}_{m})}|} \\
        NMAE_a = \frac{1}{3n}\sum_{m}^3\sum_{i}^{n}{|\frac{p_{i,m} - y_{i,m}}{max(\mathbf{p}_{m})-min(\mathbf{p}_{m})}|} 
    \end{gather}
    \item \textbf{Root mean squared error (RMSE)} calculates the squared root of the variance in the difference between prediction and ground truth.
    \begin{gather}
        RMSE_m = \frac{1}{n}\sqrt{\sum_{i}^{n}{(p_{i,m} - y_{i,m})^2}} \\
        RMSE_a = \frac{1}{3n}\sum_{m}^3\sqrt{\sum_{i}^{n}{(p_{i,m} - y_{i,m})^2}} 
    \end{gather}
    \item \textbf{Normalized Root mean squared error (NRMSE)} normalizes the RMSE normalized by the range of each dimension.
    \begin{gather}
        NRMSE_m = \frac{\sqrt{\sum_{i}^{n}{(p_{i,m} - y_{i,m})^2}}}{n(max(\mathbf{p}_{m})-min(\mathbf{p}_{m}))} \\
        NRMSE_a = \frac{1}{3n}\sum_{m}^3\frac{\sqrt{\sum_{i}^{n}{(p_{i,m} - y_{i,m})^2}}}{max(\mathbf{p}_{m})-min(\mathbf{p}_{m})} 
    \end{gather}
\end{enumerate}
Here $n$ represents the total number of users in the test labeled dataset, $m\in\{x,y,z\}$ indicates the single dimension, and $a$ denotes the metric averaged across the dimension. These metrics offer a comprehensive assessment of model performance in user localization.

\begin{table*}[!htp]
\begin{center}
\caption{Numerical results}
\label{tab:perf}\begin{tabular}{c|cc|cccc|cccc}
\hline
\multirow{2}*{Model} & \multicolumn{2}{c|}{Implementation} & 
\multicolumn{4}{c|}{MAE} & 
\multicolumn{4}{c}{NMAE} \cr
\cline{2-11} & Pretraining & Finetuning & x-axis & y-axis& z-axis& average &x-axis & y-axis& z-axis& average\cr 
\hline 
Model 1 & \multicolumn{2}{c|}{MLP} & 46.0527	& 73.8419	&62.8147&	60.9031&	0.0713&	0.0783&	1.5371&	0.5622 \cr
Model 2 & \multicolumn{2}{c|}{CNN}& 26.2827&	30.5357&	\textbf{8.1639}&	21.6608	&0.0407&	0.0324&	\textbf{0.1998}&	0.0909
\cr \hline 
Model 3 & MLP & MLP &38.3670 &	50.6702 &	8.4476	&32.4949&	0.0594	& 0.0537	&0.2067	&0.1066 \cr
Model 4 & CNN & MLP &\textbf{18.4095}&	\textbf{21.2148}	&10.9803	&\textbf{16.8682}&	\textbf{0.0285}	&\textbf{0.0225}	&0.2687&	\textbf{0.1065} \cr
\hline
\end{tabular}
\begin{tabular}{c|cc|cccc|cccc}
\hline
\multirow{2}*{Model} & \multicolumn{2}{c|}{Implementation} & 
\multicolumn{4}{c|}{RMSE} & 
\multicolumn{4}{c}{NRMSE} \cr
\cline{2-11} & Pretraining & Finetuning & x-axis & y-axis& z-axis& average& x-axis & y-axis& z-axis& average\cr 
\hline 
Model 1 & \multicolumn{2}{c|}{MLP} & 63.6843& 	106.6284	& 91.1460& 	89.0806& 	0.0985	& 0.1131& 	2.2303	& 0.814  \cr
Model 2 & \multicolumn{2}{c|}{CNN} &35.0762	&44.4942&	10.8762&	33.3804&	0.0543&	0.0472	&0.2661	&\textbf{0.1225} \cr \hline 
Model 3 & MLP & MLP & 54.1198& 	81.9006	& \textbf{10.0560}& 	57.1584	& 0.0837& 	0.0869& 	\textbf{0.2461}	& 0.1389 \cr
Model 4 & CNN & MLP &\textbf{29.4605}	& \textbf{31.4541}& 	12.9957	& \textbf{26.1507}	& \textbf{0.0456}	& \textbf{0.0334}& 	0.3180& 	0.1323 \cr\hline
\end{tabular}
\end{center}
\end{table*}

\section{Results and Discussion}
The quantitative results of our model are presented in Table \ref{tab:perf}. 
% We build 4 different model, Model 1 and 2 do not use any pretraining and are based on supervised learning to predict the user's location. Model 3 and 4 use self supervised based pretraining and are then use MLP finetuning to make the final predictions. The table shows the Mean Absolute Error (MAE), Normalized MAE, Root Mean Squared Error (RMSE) and Normalized RMSE. (repeated, already described previously)
%From the table, several noteworthy observations emerge first.
The table yields several notable insights:
(1) Models employing supervised learning only (Model 1, 2) consistently underperform those self-supervised based pretrained models (Model 3, 4). 
(2) Across different implementations, CNN-based models consistently outperform MLP-based models.

In a direct comparison between Model 1 and Model 2, 
(1) the CNN-based model (Model 2) outperforms MLP based on Model 1 by a significant margin. 
% Specifically, the average MAE values for the $x$, $y$, and $z$ coordinates are 26.2827, 30.5357, and 8.1639 meters respectively for Model 2, whereas they are 46.0527, 73.8419, and 62.8147 meters respectively for Model 1. 
Notably, MAE for $x$ and $y$ in Model 1 are more than twice as high as those in Model 2, while for $z$ axis, it is as much as 7 times higher. On average, the MAE for Model 1 (60.9031) exhibits an MAE approximately 3 times that of Model 2 (21.6608). (2) A similar trend is observed in the NMAE values as well, with Model 1 displaying values for $x$ and $y$ axis approximately six times higher on average than Model 2; for $z$ it is about 7 times higher. The NMAE value averaged across three dimensions for Model 2 is almost 6 times higher than that of Model 1.
(3) The RMSE values are also consistently much higher for Model 1 compared to Model 2, with the average RMSE of Model 1 being almost 2.6 times higher than that of Model 2 (89.0806 compared to 33.3804).

In examining Model 3 and Model 4, both models utilize pretrained frameworks to enhance user localization. The following observations are made:
(1) 
%However it is interesting to note that Model 3, which used an MLP-based self-supervised training model, does not perform better than Model 2, which uses a CNN-based model without pretraining. A closer comparison reveals that MAE for $x$ and $y$ are higher for Model 3 than for Model 2, and while Model 3 performs better than Model 1 for $z$ axis, it still lags behind Model 2. This can be attributed to the fact that CNNs can capture spatial relationships quite well compared to MLPs. 
Interestingly, Model 3, an MLP-based self-supervised model, does not exhibit better performance compared with Model 2, which utilizes a CNN-based model without pretraining. A detailed comparison shows that MAE for the $x$ and $y$ axes are higher for Model 3 than for Model 2, and while Model 3 fares better than Model 1 for the $z$ axis, it still falls short compared with Model 2. This trend suggests that CNNs can better capture spatial relationships compared to MLPs.
(2) In contrast, Model 4 overcomes this limitation by employing a CNN-based AE model. Model 4 achieves the minimal MAE for $x$ axis (18.4095) and $y$ axis (21.2148) and consequently the minimum average MAE of all the models (16.8682 meters).
(3) Notably, Model 4 records a higher MAE for $z$ axis compared with both Model 3 and Model 2, potentially indicating overfitting, even though we did not encounter such issues during training. Simpler models perform more effectively on $z$ axis. 
(4) In terms of RMSE and NRMSE, Model 4 clearly outperforms all other models except for MAE on $z$ axis, but it is able to achieve significantly lower average values across all the models.

\section{Acknowledgment}
This work was supported in part by the Federal Highway Administration (FHWA) Exploratory Advanced Research (EAR)
under Grant 693JJ320C000021.
\section{Conclusions}
This paper has effectively highlighted the potential of self-supervised learning in enhancing supervised learning performance for user localization using CSI data. We designed four distinct models: two utilized only supervised learning with labeled data, while the other two leveraged self-supervised pretraining with unlabeled data, followed by supervised finetuning with labeled data. Remarkably, our findings indicate the superior performance of the CNN-based pertaining model with an average MAE of 16.8682 meters, surpassing all other models by a considerable margin. Additionally, our research underscores the suitability of CNNs over MLPs for both Self-Supervised and Supervised Learning in this context. Furthermore, we demonstrate that leveraging unlabeled data through Self-Supervised Learning can effectively facilitate user localization when dealing with large geographical areas.

\bibliographystyle{IEEEtran}
\bibliography{references}
\end{document}